\documentstyle[12pt]{article}
\newcommand{\II}{{\cal I}}
\newcommand{\MM}{{\cal M}}
\newcommand{\GG}{{\cal G}}
\newcommand{\JJ}{{\cal J}}

\newcommand{\be}{\begin{equation}}
\newcommand{\ee}{\end{equation}}
\newcommand{\ben}{\begin{eqnarray}\displaystyle}
\newcommand{\een}{\end{eqnarray}}
\newcommand{\refb}[1]{(\ref{#1})}

\newcommand{\sectiono}[1]{\section{#1}\setcounter{equation}{0}}

\begin{document}

{}~ \hfill\vbox{\hbox{hep-th/9609176}\hbox{MRI-PHY/96-28}}\break

\vskip 3.5cm

\centerline{\large \bf Unification of String Dualities}

\vspace*{6.0ex}

\centerline{\large \rm Ashoke Sen\footnote{On leave of absence from 
Tata Institute of Fundamental Research, Homi Bhabha Road, 
Bombay 400005, INDIA}
\footnote{E-mail: sen@mri.ernet.in, sen@theory.tifr.res.in}}

\vspace*{1.5ex}

\centerline{\large \it Mehta Research Institute of Mathematics}
 \centerline{\large \it and Mathematical Physics}

\centerline{\large \it 10 Kasturba Gandhi Marg, 
Allahabad 211002, INDIA}

\vspace*{4.5ex}

\centerline {\bf Abstract}

We argue that all conjectured dualities involving various string,
$M$- and $F$- theory compactifications can be `derived' from the conjectured
duality between type I  and SO(32) heterotic string theory,  T-dualities,
and the definition of $M$- and $F$- theories. 

(Based on a talk given at the conference on `Advanced Quantum Field Theory',
Le Londe-les-Maures, France, Aug.31-Sept.5, 1996, in memory of Claude
Itzykson)

\vfill \eject

\baselineskip=18pt

\sectiono{Introduction} \label{s1}

During the last two years, there has been an explosion of conjectures
describing new relations between apparently unrelated string 
theories\cite{SREV,SCHREV,POLREV,DINREV}. 
The purpose of this article will be not to add new conjectures to this
list, but to provide a link between these various duality conjectures. In
particular I shall try to identify a minimal set of duality conjectures
from which all other duality conjectures can be `derived'.

There are five known string theories in ten dimensions which have been
named type IIA, type IIB, type I, SO(32) heterotic and E$_8\times$E$_8$
heterotic. In order to relate these five theories,
we need at least four relations.
One of them comes from the observation\cite{SAGN,HORA,POLET,GIMPOL}
that the type I
theory can be regarded as the quotient of the type IIB theory by the
world sheet parity transformation $\Omega$. Two more duality relations
follow from the T-duality relations that can be verified order by order
in perturbation theory. These can be stated as follows:
\begin{itemize}
\item{
Type IIA theory compactified on a circle of radius $R$ is dual to type
IIB theory compactified on a circle of radius $1/R$\cite{IIAIIB,POLET}.}
\item{SO(32) heterotic theory compactified on a circle of radius $R$, with
gauge group broken down to SO(16)$\times$SO(16) by Wilson
line is dual to E$_8\times$E$_8$ heterotic theory compactified on a circle
of radius $1/R$ with gauge group broken down to SO(16)$\times$SO(16)
by Wilson line\cite{NARAIN,SOE8}.}
\end{itemize}
Using these three relations we can divide the five string theories into
two families, the type II family which includes the two type II and the type
I theories, and the heterotic family that includes the two heterotic
string theories. The members within each family are related to each other
by T-duality transformations or quotienting. Thus in order to relate all five
string theories, we need at least one
more relation between these two families.
This is provided by the type I - SO(32) duality conjecture which states
that\cite{WITTEND}
\begin{itemize} 
\item{In ten dimensions type I theory with coupling constant $g$ 
is dual to SO(32) heterotic string theory with coupling constant $1/g$.}
\end{itemize}
We shall see that all other conjectured dualities among various string
theories and their compactifications can be `derived' from these four
relations between the five string theories.

Before I go on, I would like to make two cautionary remarks. 
First of all, we should recall that since duality in general relates a
weakly coupled string theory to a strongly coupled string theory, any
`derivation' of a duality requires manipulations involving strongly coupled
string theory. In the absence of a non-perturbative formulation of string
theory, the rules that we must follow while carrying out these 
manipulations are not known {\it a priori} and need to be discovered
on the way. We shall see that there are a reasonable set
of rules that one can follow during these manipulations, using which
we can `derive' all known string dualities from a few simple ones.

The second remark that I would like to make is that one should not 
interprete the result of this paper
by regarding these `parent' dualities as more
fundamental than any other duality. One of the main lessons that we 
have learned during the last two years is that no string theory is more
fundamental than any other, although some theory  might be
more suitable than
others for describing a specific region of the moduli space. I believe
the same is true for string dualities, since there are complicated
interconnections between various dualities. Nevertheless it might be useful
to identify a minimal set of dualities from which others can be `derived'.

The paper will be organized as follows. In section 2 I shall outline
the set of rules that we shall be using for `deriving' one duality
relation from another. In section 3 we shall see how these rules can be
used to derive all conjectured dualities involving conventional string
compactification by starting from the SO(32) heterotic - type I duality
in ten dimensions. In particular, I shall discuss in some detail the
`derivation' of the six dimensional string-string duality between heterotic
string theory on $T^4$ and type IIA string theory on K3, and also of the
SL(2,Z) self-duality of type IIB string theory in ten dimensions. Sections
4 and 5 will be devoted to dualities involving compactification of $M$-
and $F$-theories respectively. During our analysis in sections 3-5 we
shall be making use of T-duality symmetries indiscriminately, but in
the appendix we shall identify the minimal set of T-dualities from which
all other T-duality symmetries can be derived. 

\sectiono{Rules for Derivation} \label{s2}

In this section I shall review the various rules that we shall be
following in order to `derive' new duality relations among string
theories from a given duality relation. I should mention here that
most of the manipulations discussed in this section are valid in the
form stated only for theories for which the local 
structure of the moduli
space is not corrected by quantum effects. This is automatic for
theories with sixteen or more supercharges since for these theories
supersymmetry determines the local structure of the moduli space
completely. Since the detailed application of these methods, 
discussed in the next three sections, involve theories of this kind
only, we are on a firm ground as far as these examples are concerned.
For theories with less number of supersymmetries the moduli space
gets quantum corrected, and hence even the T-duality symmetries are
modified by these corrections\cite{DEWLUS,ANTNAR}. We do expect however
that the methods of this section can still be used to determine
dualities between such theories, although we shall not be able to
reproduce the precise map between the quantum corrected moduli
spaces of the two theories by these methods. A notable example of
this is the `derivation' of the duality between type IIA theory on
a Calabi-Yau manifold and heterotic string theory on $K3\times T^2$.
We shall discuss this briefly in the next section.

\subsection{T-Dualities} \label{s2.5}

T-dualities are the best understood duality symmetries in string theory.
They hold order by order in string perturbation
theory to all orders, and even though they have not been established for
full non-perturbative string theory,\footnote{This point 
has been emphasized in ref.\cite{PORR}.}
we shall
use them indiscriminately during our analysis by assuming that they
are symmetries of the full non-perturbative string theory. 

It is nevertheless useful to identify a minimal set of T-dualities
from which others can be derived.
Part of the T-duality symmetries in
various string theories can be identified as the global diffeomorphism
group of the manifold on which the string theory has been compactified.
This is part of the general coordinate transformation in the 
corresponding string theory before compactification.
Similarly part of the T-duality symmetries
might be associated with Yang-Mills gauge symmetries of string theory.
Validity of these T-duality symmetries in the full non-perturbative string
theory is therefore on a firm footing.
In the appendix we shall identify the minimal set of T-duality
conjectures which need to be added to these gauge symmetries
in order to derive the full
set of T-duality relations in string theory.

\subsection{Unifying $S$- and $T$-Dualities} \label{s2.2}
Suppose a theory $A$ compactified on a manifold $K_A$ 
(this might represent a string, $M$- or $F$-theory compactification) 
is known to have a
self-duality group $G$. Let us now further compactify this theory on another
manifold $\MM$, and suppose that this theory has a T-duality group $H$.
It is reasonable to assume $-$ and we shall assume $-$
that this theory is also invariant under the
duality symmetry group $G$ of the theory before compactification on $\MM$.
Quite often one finds that the elements of $G$ and $H$ don't commute
with each other and together generate a much larger group $\GG$. {}From
this we can conclude that the full duality symmetry group of theory $A$
compactified on $K_A\times \MM$ is at least $\GG$. This method has been
successfully used to derive the duality group
in many cases. For example, it was used to
show\cite{SENTHREED} that the full duality symmetry group 
of heterotic string theory
compactified on $T^7$ is O(8,24;Z) which includes the T-duality group
O(7,23;Z) of this compactification, as well as the SL(2,Z) S-duality
group of heterotic string theory compactified on $T^6$. This method has
also been used to derive the duality groups of type II theories compactified
on $T^n$ from the knowledge of T-duality group of the corresponding
compactification, and the SL(2,Z) S-duality group of type IIB theory in
ten dimensions\cite{HULTOW}. 
For example type IIB theory compactified on $T^4$ has
a T-duality group SO(4,4;Z). These SO(4,4;Z) transformations do not in
general commute with the SL(2,Z) S-duality transformation of the type
IIB theory, and together they generate a duality symmetry group 
SO(5,5;Z) which has been conjectured to be the full duality
symmetry group in this theory.

\subsection{Duality of Dualities} \label{s2.3}
This is a generalization of the procedure described in the previous
subsection to the case of dualities relating two different theories.
Suppose theory $A$ on $K_A$ is dual to theory $B$ on $K_B$. Let us now 
compactify both theories further on a smooth manifold $\MM$.
In the spirit of the previous subsection we shall assume that these
two theories are still dual to each other.
Let $H_A$ be the
$T$-duality group of the first theory and $H_B$ be the T-duality group
of the second theory. Typically $H_A$ and $H_B$ are not isomorphic;
in fact quite often the image of $H_B$ in the first theory involves 
transformations which act non-trivially on the coupling
constant. The same is true for the image of $H_A$ in the second theory.
Thus $H_B$ ($H_A$) in general contains information about
non-perturbative duality symmetries in the first (second) theory. In
particular, the
full duality group of the first theory must include $H_A$, as well 
as the image of $H_B$. 
By studying the full group generated by
$H_A$ and the image of $H_B$ we can quite often determine the full duality
group of the theory. 

The most well known example of this kind is the `derivation' of $S$-duality
of heterotic string theory on $T^6$ from string-string duality in six
dimensions\cite{DUFDD,WITTEND}. 
We start from the duality between type IIA on K3 and heterotic
on $T^4$\cite{HULTOW}, 
and compactify both theories on $T^2$. The first theory has
an SL(2,Z) T-duality symmetry associated with the torus compactification
of the type IIA theory. Under string-string duality this transformation
gets mapped to the S-duality of the heterotic string theory on $T^6$. Thus
the six dimensional string-string duality conjecture together with the
T-duality of type IIA on $T^2$ implies S-duality of heterotic string theory
on $T^6$.

\subsection{Taking Large Size Limit of Compact Manifolds} \label{s2.4}

Suppose a theory $A$ on $K_A$ is known to have a self duality group
$\GG_A$. Let us now try to recover the theory $A$ in flat space-time
by taking the large size limit of the manifold $K_A$. In general, the
duality group $\GG_A$ will not commute with this limit, since a typical
element of the duality group will map a large size $K_A$ to a small
or finite size $K_A$. Thus not all of $\GG_A$ will appear as duality
symmetry of the theory $A$ in flat space-time. However suppose $G_A$
is the subgroup of $\GG_A$ that commutes with this limit; {\it i.e.}
elements of $G_A$, acting on a large size $K_A$, produces a large size
$K_A$. In that case we can conclude that $G_A$ is the duality symmetry
group of the theory $A$ in flat space-time\cite{ASPMOR,DUFTRI,HULTWE}. 
(Sometime, if the duality
group of the compactified theory contains global diffeomorphism
symmetries of the compact manifold $K_A$, then in the large volume
limit it becomes part of the general coordinate transformation of
the non-compact manifold, and does not give any new information).

I shall illustrate this idea through an example. Let us consider
type IIA and type IIB theories compactified on a circle. These two
theories are related by T-duality, and both have
duality group SL(2,Z) which act on the moduli fields as 
\be \label{en1}
\lambda_{IIB}\to {p\lambda_{IIB} + q\over r\lambda_{IIB} + s}, \qquad 
\Psi\to \Psi\, ,
\ee
where $p,q,r,s$ are integers satisfying $ps-qr=1$, $\lambda_{IIB}$ is a
complex scalar and $\Psi$ is a real scalar field. In terms of variables
of the type IIB theory, $\lambda_{IIB}$ and $\Psi$ are given by
\be \label{en2}
\lambda_{IIB}=a_{IIB} + i e^{-\Phi_{IIB}/2}, \qquad \Psi = \Phi_{IIB}
-8\ln R_{IIB}\, ,
\ee
where $a_{IIB}$ denotes the scalar arising in the Ramond-Ramond 
(RR) sector, 
$\Phi_{IIB}$ denotes
the ten dimensional dilaton, and $R_{IIB}$ is the radius of the circle 
measured in the type IIB metric. On the other hand, in terms of the
type IIA variables, we have
\be \label{en3}
\lambda_{IIB}=(A_{IIA})_9 + i e^{-\Phi_{IIA}/2 + \ln R_{IIA}}, \qquad 
\Psi = \Phi_{IIA} + 6\ln R_{IIA}\, ,
\ee
where $(A_{IIA})_\mu$ denotes the $\mu$-th component of the RR vector field,
$x^9$ denotes the direction of the circle, $\Phi_{IIA}$ is the ten
dimensional dilaton of the type IIA theory and $R_{IIA}$ is the
radius of the circle measured in the type IIA metric.

Let us now try to recover the ten dimensional type IIB theory by taking
the limit $R_{IIB}\to\infty$ keeping $\Phi_{IIB}$ and $a_{IIB}$
fixed. As is clear from \refb{en2}, in this limit 
$\lambda_{IIB}$ is fixed, and
$\Psi\to -\infty$. The SL(2,Z) duality transformations described in
eq.\refb{en1} does not affect this limit. Thus we can conclude from
this that the SL(2,Z) duality of type IIB theory on $S^1$ implies
SL(2,Z) duality of the ten dimensional type IIB theory.

Let us now try to do the same thing for the type IIA theory by taking the
limit $R_{IIA}\to\infty$ keeping $\Phi_{IIA}$ fixed. We see from 
eq.\refb{en3} that in this limit $Im(\lambda_{IIB})\to\infty$. 
But an SL(2,Z) transformation of the form given in eq.\refb{en1}
does not preserve this condition; in fact acting on a configuration with
large $Im(\lambda_{IIB})$ an SL(2,Z) transformation with $r\ne 0$ will
produce a configuration with small $Im(\lambda_{IIB})$. Thus we see that
in this case SL(2,Z) duality of type IIA theory compactified on
a circle cannot be used to conclude that the type IIA theory in
ten dimensions has an SL(2,Z) duality symmetry.

A variant of this idea can also be used for deriving dual pairs of
theories. Suppose theory $A$ on $K_A$ is dual to theory $B$ on $K_B$.
If we now take the large size limit of $K_A$, then typically it will
correspond to small or finite size of $K_B$. However if it so happens that
in the limit when the size of $K_A$ approaches infinity, the size of
$K_B$ approaches infinity as well, with the coupling constant remaining
finite, then we can conclude that the two
theories $A$ and $B$ are dual to each other in flat space-time.

\subsection{Fiberwise Application of Duality Transformation} \label{s2.1}

Suppose we are given that a theory $A$  on a compact manifold $K_A$
is dual to another theory $B$ on a compact manifold $K_B$. (More precisely
the theory $A$ on $K_A\times R^{n,1}$ is dual to theory $B$ on $K_B\times
R^{n,1}$, where $R^{n,1}$ is a Minkowski space of signature $(n,1)$.)
This duality
relation automatically comes with a map between the moduli spaces
of the two theories, with the moduli specifying
the geometry of $K_A$ ($K_B$) as well
as various background field configurations in the two theories. Let us
now consider compactification of the theories $A$ and $B$ on
two new manifolds $E_A$ and $E_B$ respectively, obtained by fibering
$K_A$ and $K_B$ on another manifold $\MM$. This means that $E_A$ ($E_B$)
is obtained by erecting at every point in $\MM$ a copy of $K_A$ ($K_B$),
with the provision that the moduli of $K_A$ ($K_B$) 
(including possible background fields that arise upon
compactifying $A$ ($B$) on $K_A$ ($K_B$)) could
vary as we move on $\MM$.  Thus if $\vec y$ denote the coordinates on
$\MM$, and $\vec m_A$ and $\vec m_B$ denote the coordinates on the moduli
spaces of theory $A$ on $K_A$ and theory $B$ on $K_B$ respectively, then
specific compactification of $A$  and $B$ on
$E_A$ and $E_B$ are specified by the functions $\vec m_A(\vec y)$ and
$\vec m_B(\vec y)$ respectively. The original duality between $A$ on $K_A$
and $B$ on $K_B$ then gives rise to a map between the moduli spaces
of $A$ on $E_A$ and $B$ on $E_B$ as follows: 
given a function $\vec m_A(\vec y)$, we
construct $\vec m_B(\vec y)$ by applying the original duality map at
every point $\vec y$ on $\MM$.

It has been argued by Vafa and Witten\cite{VAFWIT} 
that the theory $A$ on $E_A$ is dual
to the theory $B$ on $E_B$ under this map. The basis of this
argument is as follows.  
As long as the moduli of $K_A$ and $K_B$ vary slowly
on $\MM$, near a local neighbourhood of any point $P$ on $\MM$ the two
manifolds effectively look like $K_A\times R^{n,1}$ and $K_B\times R^{n,1}$
respectively.  Thus we can
relate the two theories by applying the original duality between $A$ on
$K_A\times R^{n,1}$ and $B$ on $K_B \times R^{n,1}$. This
argument breaks down when the moduli of $K_A$ and $K_B$ vary rapidly on
$\MM$, in particular near singular points on $\MM$ where the fiber 
degenerates. However, in many examples that have been studied, the duality
between $A$ on $E_A$ and $B$ on $E_B$ continues to hold even in the
presence of such singular points. Perhaps the lesson to be learned from
here is that the presence of singular points of `measure zero' does not affect
the duality between the two theories. Put another way, the fact that in the
bulk of the manifolds $E_A$ and $E_B$ the two theories are equivalent by
the original duality relation, forces them to be equivalent even on 
these singular subspaces of codimension $\ge 1$, even though the original
argument breaks down on these subspaces. In any case, we shall henceforth
assume that for $E_A$ and $E_B$ constructed this way, $A$ on $E_A$ is dual
to $B$ on $E_B$.

This method was used to `derive' the conjectured 
dualities\cite{KACVAF,KLM,ALDFON,SCHLYN} 
between type
IIA string theory on a Calabi-Yau manifold and the heterotic 
string theory
on $K3\times T^2$ from the string-string duality conjecture
in six dimensions that
relates type IIA on $K3$ and heterotic on $T^4$. This is done by 
representing the Calabi-Yau manifold as $K3$ fibered over $CP^1$ and 
the corresponding $K3\times T^2$ as $T^4$ fibered over $CP^1$.

A special case of this is a pair of $Z_2$ orbifolds constructed as follows.
Suppose we have a dual pair of theories 
($A$ on $K_A$) and ($B$ on $K_B$). Let
us compactify both theories further on a smooth manifold $\MM$, and let us
assume that both these theories have a $Z_2$ symmetry group that are
related to each other under the duality map. We can divide the action of the
$Z_2$ transformation into two parts: let $s$ represent the geometric action
on the manifold $\MM$ which is identical in the 
two theories, and $h_A$ ($h_B$)
denote the geometric action on the manifold $K_A$ ($K_B$) as well as internal
symmetry transformation in theory $A$ ($B$). We now compare the two
quotient theories 
\centerline{
($A$ on $K_A\times \MM/h_A\cdot s$) and 
($B$ on $K_B\times \MM/h_B\cdot s$)  
}
where by an abuse of notation we have denoted the $Z_2$ group by its
generator $h_A\cdot s$ ($h_B\cdot s$).
A little bit of mental exercise shows us that the first theory has the
structure of $A$ compactified
on a fibered space $E_A$ with base $\MM/s$, and fiber $K_A$,
with twist $h_A$ on the fiber as we go from any point $P$ 
on $\MM$ to $s(P)$.
(Note that this is a closed cycle on $\MM/s$.) 
Similarly the second theory
has the structure of $B$ compactified
on $E_B$, where $E_B$ is a fibered space with
base $\MM/s$ and fiber $K_B$, $h_B$ being the twist on the
fiber as we go from the
point $P$ to $s(P)$ on $\MM$. Thus our previous argument will tell us that
these two theories are dual to each other. Note that if $P_0$ denotes a
fixed point of $s$, then at $P_0$ the fiber degenerates to 
$K_A/h_A$ ($K_B/h_B$), and the original duality between $A$ on $K_A$
and $B$ on $K_B$ fails to relate these two quotient theories. 
But as has been stated before, as long as these
are isolated fixed points (hyperplanes) of codimension $\ge 1$ we expect the
duality between the two resulting theories to hold. This method was used in
ref.\cite{FHSV} to construct a dual pair of string theories in four
dimensions with N=2 supersymmetry.

The two extreme cases of this construction are:
\begin{enumerate}
\item{$s$ acts on $\MM$ without fixed points. In this case $\MM/s$ has no
fixed points and the fiber never degenerates. The case for duality
between the two resulting theories is on a firmer footing in this case.}
\item{$s$ acts trivially on $\MM$. In other words $s$ is just the
identity transformation. In this case the fiber is everywhere $K_A/h_A$
($K_B/h_B$) and there is no reason to expect the resulting theories to be
dual to each other. In fact there are many known examples of this kind
where the resulting theories are {\it not} dual to each other.}
\end{enumerate}

In principle the above construction should extend to more general orbifolds,
but there are interesting subtleties involved in such 
extensions\cite{BLUZAF,DABPAR3,GOPMUK,BLUM}.

\sectiono{Minimal Set of String Dualities} \label{s3}

I begin by briefly
reviewing various known duality conjectures involving conventional
string compactifications. In ten
dimensions there are two non-trivial duality conjectures: the SL(2,Z)
self-duality of type IIB sting theory and the duality between type I
and SO(32) heterotic string theories. Upon compactifying type IIB or type
IIA theory on an $n$ dimensional torus, we get a large duality
symmetry group, but it can be derived by combining the T-duality
symmetries of type IIA/IIB string theories and the SL(2,Z) self-duality
of the ten dimensional type IIB string theory\cite{HULTOW}.
As we go down in dimensions, we encounter the next non-trivial duality
conjecture in six dimensions: the duality between type IIA string theory 
compactified on $K3$ and heterotic string theory compactified on $T^6$,
also known as the string-string duality conjecture\cite{HULTOW}. 
As has already been
pointed out, staring from this duality conjecture, we can `derive' the
$S$-duality of heterotic string theory on $T^6$ by mapping it to a
$T$-duality symmetry of type IIA on $K3\times T^2$. Furthermore, all
the conjectured dualities involving type IIA on Calabi-Yau manifolds
and heterotic string theory on $K3\times T^2$ can be recovered from
the string-string duality conjecture by representing the Calabi-Yau 
manifold as $K3$ fibered over $CP^1$ and $K3\times T^2$ as $T^4$
fibered over $CP^1$\cite{KACVAF,KLM,ALDFON,SCHLYN}.
Finally there are strong-weak coupling duality conjectures involving
compactification of $E_8\times E_8$ 
heterotic string theory on K3\cite{DUFMINW}. By using a T-duality 
transformation one can relate this theory to SO(32) heterotic
string theory on K3, and
the type I - SO(32) heterotic duality in ten dimension relates this
further to an orientifold compactification\cite{SCHET}. Under this
duality, the strong-weak coupling duality in the original
theory can be mapped to a
T-duality transformation in the orientifold theory\cite{SCHET}. Thus
the strong-weak coupling duality of $E_8\times E_8$
heterotic string theory on K3 can also
be `derived' in terms of other duality conjectures.

Thus at this stage it would appear that there are three independent
strong-weak coupling duality conjectures involving string 
compactification:
\begin{enumerate}
\item{Duality between type I and SO(32) heterotic string theories
in ten dimensions.}
\item{SL(2,Z) self-duality of type IIB string theory in ten dimensions.}
\item{Duality between type IIA string theory on K3 and heterotic 
string theory on $T^4$.}
\end{enumerate}
There are many other duality conjectures involving string 
compactifications, but all of them can be `derived' from these 
using the procedures outlined in section \ref{s2}. 

We shall now see that
\begin{itemize}
\item{(3) can be `derived' from (1) and (2), and}
\item{(2) can be `derived' from (1),}
\end{itemize}
using the rules given in section \ref{s2}.
Thus the only independent non-perturbative duality conjecture is that
between type I and SO(32) heterotic string theory in ten dimensions. 

\subsection{Derivation of String-String Duality} \label{s3.1}

The analysis in this subsection will follow closely that in
ref.\cite{DUORBI}.
I begin with a review of the symmetries of
type IIB string theory in ten dimensions. This theory
has two global $Z_2$ symmetries which hold order by order in string
perturbation theory. The first one involves changing the sign of all
Ramond sector states on the left, and is denoted by $(-1)^{F_L}$. Acting
on the bosonic fields of the theory it changes the sign of
all the fields coming from the RR sector, leaving all
the fields in the Neveu-Schwarz-Neveu-Schwarz (NS) sector invariant.
The second $Z_2$ symmetry involves a world-sheet parity transformation,
and is denoted by $\Omega$. It changes the sign of the anti-symmetric rank
two tensor field $B_{\mu\nu}$ coming from the NS sector, and the scalar
$a_{IIB}$ and the rank four anti-symmetric tensor 
field $D_{\mu\nu\rho\sigma}$
coming from the RR sector, leaving all the other
massless bosonic fields invariant.

Besides these perturbative duality symmetries, the ten dimensional type
IIB theory also has a conjectured SL(2,Z) symmetry. This transformation
leaves the canonical metric invariant and acts on the other massless
bosonic fields in the theory as:
\be \label{e1}
\lambda_{IIB}\to {p\lambda_{IIB} + q\over r\lambda_{IIB} + s}, \qquad
\pmatrix{B'_{\mu\nu}\cr B_{\mu\nu}} \to \pmatrix{p & q\cr r & s}
\pmatrix{B'_{\mu\nu}\cr B_{\mu\nu}}\, ,
\ee
where $p,q,r,s$ are integers satisfying
\be \label{e1a}
ps-qr=1\, .
\ee
$B'_{\mu\nu}$ is the rank two antisymmetric tensor field arising in
the RR sector of the theory, and $\lambda_{IIB}$ has been defined in
eq.\refb{en2}.
The special SL(2,Z) transformation generated by
\be \label{e3}
\pmatrix{p & q\cr r & s} = \pmatrix{ 0 & 1\cr -1 & 0}\, ,
\ee
is denoted by $S$.

Studying the action of the transformations $(-1)^{F_L}$, $\Omega$ and
$S$ on the various massless fields in this theory one can show that
\be \label{e4}
S (-1)^{F_L} S^{-1} = \Omega\, .
\ee
We shall now construct a new dual pair of theories following the
orbifolding procedure described in section \ref{s2}. 
According to this procedure, if theory $A$ on $K_A$ is dual to theory
$B$ on $K_B$, then $A$ on $K_A\times \MM/h_A\cdot s$ is dual to
$B$ on $K_B\times \MM/h_B\cdot s$. We shall choose $A$ on $K_A$
to be the type IIB
theory, $B$ on $K_B$ to be the $S$-transformed type 
IIB theory, $h_A$ to be
$(-1)^{F_L}$, $h_B$ (the image of $h_A$ under $S$) to be $\Omega$,
$\MM$ to be a four dimensional torus $T^4$ labelled by $x^m$
($6\le m\le 9$) and $s$ to be the transformation $\II_4$ that
transforms $x^m$ to $-x^m$ for
$(6\le m\le 9$). Thus we arrive at the duality:
\be \label{e5}
\hbox{(type IIB on $T^4/(-1)^{F_L}\cdot\II_4$)} \leftrightarrow
\hbox{(type IIB on $T^4/\Omega\cdot\II_4$)}\, .
\ee
Let us now perform an $R\to (1/R)$ duality transformation on the $x^6$ 
coordinate in the first theory. This is a $T$-duality
transformation that converts the type IIB theory to type IIA theory. It
also maps the transformation $(-1)^{F_L}\cdot\II_4$ in the type IIB
theory to the transformation $\II_4$ in the type IIA theory $-$ this can be
checked either by studying the action of various transformations on the
massless fields of the theory, or by studying their action on the
world-sheet fields. Thus the
first theory is $T$-dual to type IIA on $(T^4)'/\II_4$. This is a special
point in the moduli space of type IIA on K3.

Next let us perform $R\to(1/R)$ duality transformation 
on all the four circles 
of the second theory. This gives us back a type IIB theory, but maps the
transformation $\Omega\cdot\II_4$ to just $\Omega$. Thus the second
theory is related by a T-duality transformation to type IIB on 
$(T^4)''/\Omega$. But as mentioned in the introduction, type IIB modded
out by $\Omega$ is the type I theory. Thus the second theory is
equivalent to type I on $(T^4)''$. By the duality between type I and
SO(32) heterotic string theory in ten dimensions, this theory is
equivalent to heterotic string theory on $(T^4)''$.

Thus we have arrived at the duality between heterotic string theory on
$(T^4)''$ and type IIA string
theory on K3, {\it i.e.} the string-string duality
conjecture. Note that although this relation was `derived' at a special
point in the moduli space (orbifold limit of K3), once the exact duality
between two theories is established at one point in the moduli space, we
can go away from this special point by switching on background fields in
both theories without destroying the duality. This was done explicitly
in ref.\cite{DUORBI} for this case. In fact this analysis automatically
provides an explicit map between the moduli spaces of the two theories.

\subsection{Derivation of S-Duality of Type IIB String Theory} \label{s3.2}

The strategy that we shall follow for `deriving' 
this duality is to first study
the duality symmetries of type IIB theory compactified on an orientifold
$T^2/(-1)^{F_L}\cdot\Omega\cdot\II_2$ where 
$\II_2$ denotes the transformation
that reverses both directions on the torus, and then take the limit of large
size of $T^2$ to recover the
duality symmetries of the ten dimensional type IIB
string theory. This model was studied in detail in ref.\cite{FTHEORY}.
In this case the compact manifold $T^2/\II_2$ has the structure of a
tetrahedron, and each of the four vertices
(representing a seven dimensional hyperplane)
of the tetrahedron acts as a source of $-4$
units of RR charge associated with the nine form RR field strength. Since
$T^2/\II_2$ is a compact manifold, this
RR charge must be cancelled by putting sixteen Dirichlet 7-branes transverse
to $T^2/\II_2$. The positions of these seven branes act as extra moduli for
this compactification.

It was argued in ref.\cite{FTHEORY} that non-perturbative quantum corrections
modify the geometry of the compact manifold, and also produces a
background $\lambda_{IIB}$ field (defined in eq.\refb{en2})
that varies on the compact manifold in a
complicated manner. However, these complications do not arise in one
special case, when each vertex of the tetrahedron contains four seven branes.
In this case the geometry of the tetrahedron is flat everywhere except for
conical deficit angle $\pi$ at each of the four vertices, and the field 
$\lambda_{IIB}$ is constant on the tetrahedron. This theory has an
unbroken gauge group $SO(8)^4$, with one SO(8) factor associated with
each vertex.
For this configuration, the moduli labelling this compactification are
the volume $V_{IIB}$ of $T^2$ measured in the type IIB metric, the
complex structure modulus $\tau_{IIB}$ of $T^2$, and  $\lambda_{IIB}$.
There is no background $B_{\mu\nu}$ or $B'_{\mu\nu}$ field since they 
are odd under $(-1)^{F_L}\cdot\Omega$.

We now make the following set of duality transformations:
\begin{itemize}
\item{We first make an $R\to (1/R)$ duality transformation on both circles
of $T^2$. This takes the type IIB theory to a type IIB theory, but 
converts the transformation $(-1)^{F_L}\cdot\Omega\cdot\II_2$ to
$\Omega$\cite{DABPAR2,FTHEORY}. If we denote the new torus by $(T^2)'$,
then the resulting theory is type IIB on $(T^2)'/\Omega$, {\it i.e.}
type I theory on $(T^2)'$. Note that the gauge group is broken to SO(8)$^4$
by Wilson lines.}
\item{Using the duality between type I and SO(32) heterotic string theories,
we can now regard this theory as heterotic string theory compactified
on $(T^2)'$, with the gauge group broken to SO(8)$^4$ by Wilson lines.}
\end{itemize}
In the heterotic description we shall choose as moduli the complex
structure $\tau_{het}$ of $(T^2)'$, and,
\be \label{em1}
\rho_{het}=B^{(het)}_{89} + i V_{het}\, ,
\ee
\be \label{em2}
\Psi_{het}=\Phi_{het} - \ln V_{het}\, ,
\ee
where $B^{(het)}_{\mu\nu}$ is the rank two anti-symmetric tensor field of
the heterotic string theory, $V_{het}$ is the volume of $(T^2)'$
measured in the heterotic metric, and $\Phi_{het}$ is the dilaton of the
ten dimensional heterotic string theory. By following the duality
relations given in 
ref.\cite{WITTEND} one can find the map between the moduli in the type IIB 
description and the heterotic description. They are as follows:
\ben \label{em3}
\tau_{het} & = & \tau_{IIB}\, , \nonumber \\
\rho_{het} & = & \lambda_{IIB}\, , \nonumber \\
e^{\Psi_{het}} & = & \big(e^{-{\Phi_{IIB}\over 4}} V_{IIB}\big)^2\, .
\een

In the heterotic description the full O(18,2;Z) T-duality group of
the theory has a subgroup 
SL(2,Z)$\times$SL(2,Z)$'$ which acts on the moduli as
\be \label{em4}
\tau_{het}\to {p'\tau_{het} + q'\over r'\tau_{het} + s'}\, , \qquad
\rho_{het}\to {p\rho_{het} + q\over r\rho_{het} + s}\, , \qquad
\Psi_{het}\to \Psi_{het}\, ,
\ee
without affecting the Wilson lines.
Here $p,q,r,s,p',q',r',s'$ are integers satisfying $ps-qr=1$ and
$p's'-q'r'=1$. This implies the following duality transformations
on the type IIB moduli:
\be \label{em5}
\tau_{IIB}\to {p'\tau_{IIB} + q'\over r'\tau_{IIB} + s'}\, , \qquad
\lambda_{IIB}\to {p\lambda_{IIB} + q\over r\lambda_{IIB} + s}\, , \qquad
e^{-{\Phi_{IIB}\over 4}} V_{IIB} \to 
e^{-{\Phi_{IIB}\over 4}} V_{IIB} \, .
\ee
We now want to take the limit $V_{IIB}\to\infty$ keeping
$\tau_{IIB}$ and $\lambda_{IIB}$ fixed, and see which part of the
duality symmetries survive in this limit.\footnote{In this limit we have
SO(8) Yang-Mills gauge theory living at each of the four vertices, but
the theory in the bulk is just the ten dimensional type IIB string
theory. Thus the surviving duality group must reflect a symmetry of
the ten dimensional type IIB string theory.} {}From eq.\refb{em5} we see
that since $e^{-\Phi_{IIB}/2}\equiv Im(\lambda_{IIB})$ 
is finite before and after the transformation, 
an SL(2,Z)$\times$SL(2,Z)$'$
transformation takes large $V_{IIB}$ configuration  to large $V_{IIB}$
configuration. Thus the full SL(2,Z)$\times$SL(2,Z)$'$ duality group
survives in this limit. Of these SL(2,Z)$'$ can be identified as the global
diffeomorphism group of $T^2/\II_2$, and hence becomes part of the general 
coordinate transformation in the ten dimensional type IIB theory. The only
non-trivial duality group in this limit is then SL(2,Z), which can
easily be identified as the SL(2,Z) S-duality group of the ten dimensional
type IIB theory.

This finishes our `derivation' of the S-duality of type IIB theory from
the SO(32) heterotic - type I duality in ten dimensions, and the T-duality
symmetries. One can also find out
the SL(2,Z) transformation property of the ten dimensional metric from
this analysis. {}From eq.\refb{em5} we see that
$\exp(-\Phi_{IIB}/4)V_{IIB}$ is invariant under this
SL(2,Z) transformation. In other words, the volume of $T^2$ measured in the
metric $\exp(-\Phi_{IIB}/4)G^{(IIB)}_{\mu\nu}$ is invariant. Thus
$\exp(-\Phi_{IIB}/4)G^{(IIB)}_{\mu\nu}$ remains invariant
under the SL(2,Z) transformation.

The net result of the analysis in this section can be stated as follows:

{\it All conjectured non-perturbative
duality symmetries involving conventional string theory
compactifications can be `derived', 
according to the rules given in section \ref{s2}, from 
the duality between type I and SO(32) heterotic string theories in
ten dimensions.
}

\sectiono{Compactifications involving $M$-theory} \label{s4}

$M$-theory has been proposed as the eleven (10+1) dimensional theory which
arises in the strong coupling limit of the type IIA 
string theory\cite{TOWN,WITTEND}. The
low energy limit of $M$-theory is the 11 dimensional supergravity
theory with $N=1$ supersymmetry. More
precisely, $M$-theory compactified on a circle of radius $R$, measured
in the $M$-theory metric, is given by type IIA theory at coupling
constant
\be \label{e6}
e^{\langle\Phi_{IIA}\rangle/2}\equiv g_{IIA} = R^{3/2}\, .
\ee
The metric $G^{(M)}_{\mu\nu}$ ($0\le\mu, \nu\le 9$) of $M$-theory is
related to the metric $G^{(IIA)}_{\mu\nu}$ of the type IIA theory via
the relation
\be \label{e7}
G^{(M)}_{\mu\nu} = R^{-1}G^{(IIA)}_{\mu\nu}\, .
\ee
At present this is the only known way of defining $M$-theory. 
Thus we shall take this to be the definition of $M$-theory and use
this to `derive' other duality conjectures involving $M$-theory. 
Note however that there is a non-trivial ingredient that has gone into this
definition of $M$-theory, $-$ namely that the theory defined this way has
an eleven dimensional general coordinate
invariance. In particular, if we compactify type IIA
theory on a circle, then this would correspond to $M$-theory on a
two dimensional torus, and there should be a symmetry that exchanges
the two circles of the torus. It has been shown by Schwarz\cite{SCHWM} 
and by Aspinwall\cite{ASPIM} that this exchange symmetry can be 
traced back to the SL(2,Z)
self-duality of the ten dimensional type IIB theory. 
This SL(2,Z) transformation commutes with the decompactification
limit in which the size
of the two dimensional torus, measured in the $M$-theory metric, goes to
infinity.  Thus this
definition of $M$-theory as the strong coupling limit of type IIA
theory is consistent with eleven dimensional general coordinate 
invariance.

Let us now suppose we have $M$-theory compactified on some manifold $K$.
In order to define this theory,  
we can compactify the theory further
on a circle of radius $R$, identify this to type IIA theory on $K$, and
then take the limit $R\to\infty$
in order to recover $M$-theory on $K$.
More precisely, suppose $G^{(IIA)}_{mn}$ denotes the metric on $K$ in the
type IIA metric. Then the limit we want to take is
\be \label{e8}
\lim_{R\to\infty} \hbox{(IIA on $K$ with $g_{IIA}=R^{3/2}$, 
$G^{(IIA)}_{mn}=R G^{(M)}_{mn}$)}\, ,
\ee
with $G^{(M)}_{mn}$ fixed.

If $K$ has the structure of $S^1$ fibered over some other
manifold $\MM$, there is a direct approach to finding a dual of this
compactification. By fiberwise application of the
equivalence between $M$-theory on $S^1$ and type IIA string theory,
we can replace $M$-theory on $K$ by type
IIA on $\MM$, with the coupling constant of the type IIA theory varying
on $\MM$ according to the variation of the radius of $S^1$ on $\MM$ in the
fibration. Implicitly, this corresponds to applying the U-duality
transformation given in \refb{en1}, \refb{en3} of the type IIA theory
on $S^1$ fiberwise in the definition \refb{e8}.

All conjectured dualities involving $M$-theory can be `derived' from
these relations.
I shall illustrate this through two examples.

\subsection{$M$-theory on $T^5/Z_2$} \label{s4.1}

It has been conjectured by Dasgupta and Mukhi\cite{DASMUK} and by 
Witten\cite{WITTMD} that $M$-theory on $T^5/Z_2$ is dual to type IIB
string theory compactified 
on $K3$. Here $Z_2$ is generated by the transformation $\JJ_5$ which
changes the sign of all the five coordinates of $T^5$ and at the same
time changes the sign of the rank three anti-symmetric tensor field
$C_{MNP}$ that
appears in $M$-theory. We shall see how we can `derive' this
conjecture from the
principles outlined before. Our discussion will be brief since the
details have been given in ref.\cite{MORBI}.

By our analysis in section \ref{s2}, $T^5/Z_2$ has the structure of $S^1$
fibered over $T^4/Z_2$. Thus we can directly identify this theory to
a type IIA compactification on $T^4/Z_2$ by replacing $M$-theory on $S^1$
by type IIA theory fiberwise. By knowing the action of the
$Z_2$ transformation $\JJ_5$ on the massless 
fields of the $M$-theory, we can
find its action on the massless fields in type IIA theory. The net
result of this analysis is that it corresponds to the transformation
$(-1)^{F_L}\cdot\II_4$, where $\II_4$ denotes the reversal of sign of
all four coordinates on $T^4$. Thus $M$-theory on $T^5/\JJ_5$ is dual to
type IIA on $T^4/(-1)^{F_L}\cdot\II_4$.
An $R\to (1/R)$ duality transformation on
one of the circles of $T^4$ converts the type IIA theory to type IIB theory,
and at the same time converts the transformation $(-1)^{F_L}\cdot\II_4$
to $\II_4$. This establishes the duality between
$M$-theory on $T^5/Z_2$ and type IIB on $T^4/\II_4$,
which is simply the orbifold limit of type IIB on $K3$. 
As before, we can go away from the
orbifold limit by switching on background fields in both theories; and
the duality between these two theories at one point in the moduli space
will continue to guarantee their duality at all other points.

This principle can be used to `derive' many other duality conjectures
involving $M$-theory\cite{MORBI}. A notable example is the duality
between $M$-theory on $(K3\times S^1)/Z_2$\cite{MTHEORY} and the
Dabholkar-Park orientifold\cite{DABPAR}.

\subsection{$M$-theory on $S^1/Z_2$} \label{s4.2}

It has been conjectured by Horava and Witten\cite{HORWIT} that this
orbifold compactification of $M$-theory is dual to $E_8\times E_8$
heterotic string theory. The $Z_2$ 
transformation $\JJ_1$ changes the sign of
the coordinate labelling the circle, and also the sign of the rank
three anti-symmetric tensor field $C_{MNP}$ of $M$-theory.
If $r$ is the radius of $S^1$ and 
$g_{E_8\times E_8}$ is the coupling constant of the $E_8\times E_8$
heterotic string theory, then they are related as
\be \label{e9}
g_{E_8\times E_8}=r^{3/2}\, .
\ee
Furthermore the $M$-theory metric $G^{(M)}_{\mu\nu}$ and 
the $E_8\times E_8$
metric $G^{(E_8\times E_8)}_{\mu\nu}$ are related as
\be \label{e10}
G^{(M)}_{\mu\nu}=r^{-1} G^{(E_8\times E_8)}_{\mu\nu}\, .
\ee
We shall try to `derive' this duality from other known duality
conjectures, and the definition of $M$-theory. Since $S^1/Z_2$ does not
have the structure of $S^1$ fibered over another manifold, 
we need to use the original definition of $M$-theory compactification
as a limit of type IIA compactification. This does not require the
manifold of compactification to have $S^1$ fibration.

The manipulations that we are going to carry out are all given explicitly
in ref.\cite{HORWIT}, we shall only give a slightly different interpretation
of the results by running the argument backwards. According to eq.\refb{e8}
we define $M$-theory on $S^1/Z_2$ with radius of $S^1$
given by $r$ as
\be \label{e11}
\lim_{R\to\infty} \hbox{(IIA on $S^1/Z_2$ with $g_{IIA}=R^{3/2}$, 
$r_{IIA}=R^{1/2} r$)}\, .
\ee
By studying the action of the $Z_2$ transformation $\JJ_1$ 
on the massless fields
of the $M$-theory, one finds that in the type IIA theory it corresponds
to the transformation  $\JJ_1'$ which simultaneously changes the sign of the
coordinate labelling $S^1$ and induces a world-sheet parity 
transformation.\footnote{Note that in this case the $S^1$ of $M$-theory
compactification, on which $Z_2$ acts, represents a physical circle on 
which the type IIA theory is compactified. If on the other hand this
had represented the internal circle used in identifying $M$-theory on
$S^1$ with type IIA string theory, then the same $Z_2$ transformation 
$\JJ_1$ would have corresponded to the transformation $(-1)^{F_L}$
in type IIA theory as in the example studied in the previous subsection.}
This particular compactification of type IIA theory is known as type I$'$
theory, or more generally an orientifold\cite{SAGN,HORA,POLET,GIMPOL}. 
This has
the property that the two fixed points (which are really 8 dimensional
hyperplanes) on $S^1$, representing
boundaries of $S^1/Z_2$ if we regard $S^1/Z_2$ as a real line segment,
act as source of RR 10-form field strength. In particular,
each of these planes carry $-8$
units of RR charge. This charge needs to be cancelled by putting 16
Dirichlet 8-branes, each carrying $+1$ unit of RR charge, transverse to
$S^1/Z_2$. The positions of these $D$-branes on $S^1/Z_2$ are arbitrary, and
constitute extra moduli in this compactification of type IIA theory.

Thus we are now faced with the following question. Where shall we place
these D-branes as we take the limit \refb{e11}? Naively one would think that
if $\theta$ denotes the coordinate on $S^1$ normalized so as to have 
periodicity $2\pi$, then we place the $D$-branes at arbitrary but
fixed values of $\theta$
and then take the large radius limit of $S^1$. However, this would mean
that in the $R\to\infty$ limit the distance between the $D$-branes and
the boundaries of $S^1/Z_2$ increases. Following the analysis of 
Polchinski and Witten\cite{POLWIT} one can show that
during this process we hit singularities at finite
values of $R$ since the dilaton of the type IIA theory blows up at some
point on $S^1/Z_2$. The only way to avoid this singularity is to keep eight
of the D-branes at $\theta=0$ and eight of the D-branes at $\theta=\pi$.
In this case the dilaton is constant on $S^1/Z_2$
and we can take the $R\to\infty$ limit in \refb{e11} without encountering any
singularities. Thus we must define $M$-theory on $S^1/Z_2$ through this
limit.

For this configuration of $D$-branes the type I$'$ theory has an 
SO(16)$\times$SO(16) gauge symmetry, with the two SO(16) factors living at
the two boundaries. We shall now make the following series of duality
transformations:
\begin{enumerate}
\item{First we make an $r_{IIA}\to r_{IIA}^{-1}$ duality transformation
to map the type IIA theory into a type IIB theory. This maps the 
transformation $\JJ_1'$ into the world-sheet parity transformation $\Omega$
of the type IIB theory. The resulting theory is type IIB on $S^1$ modded out
by $\Omega$, which is just the type I theory on $S^1$. 
Since the unbroken gauge symmetry is SO(16)$\times$SO(16),
the SO(32) gauge symmetry of
type I is broken to SO(16)$\times$SO(16) by the presence of Wilson line.}
\item{Using the duality between type I and SO(32) 
heterotic string theories, 
we map this
theory to SO(32) heterotic string theory
compactified on $S^1$, with the gauge group
SO(32) broken down to SO(16)$\times$SO(16) by the presence of Wilson line.}
\item{We now make an $r_{SO(32)}\to (r_{SO(32)})^{-1}$ T-duality
transformation. This 
transforms the SO(32) heterotic string theory
to the $E_8\times E_8$ heterotic string
theory compactified on $S^1$, with the gauge group still broken to
SO(16)$\times$SO(16) due to the presence of Wilson line.}
\end{enumerate}
Throughout these set of duality transformations we can compute the coupling
constant, and the radius of $S^1$ in various theories according to the
formulae given in ref.\cite{WITTEND}. This gives the following set of
equivalent definitions of $M$-theory on $S^1/Z_2$:
\ben \label{e12}
&& \lim_{R\to\infty} \hbox{(type IIA on $S^1/Z_2$ with $g_{IIA}=R^{3/2}$, 
$r_{IIA}=R^{1/2} r$)} \nonumber \\
& \equiv & \lim_{R\to\infty} \hbox{(type I on $S^1$ with $g_{I}=Rr^{-1}$, 
$r_{I}=R^{-1/2} r^{-1}$)} \nonumber \\
& \equiv & \lim_{R\to\infty} \hbox{(SO(32) heterotic on $S^1$ with 
$g_{SO(32)}=R^{-1}r$, $r_{SO(32)}=R^{-1} r^{-1/2}$)} \nonumber \\
& \equiv & \lim_{R\to\infty} \hbox{($E_8\times E_8$ heterotic on $S^1$ with 
$g_{E_8\times E_8}=r^{3/2}$, $r_{E_8\times E_8}=R r^{1/2}$)}\, .
\een
By examining eq.\refb{e12} we see for each of the first three theories,
either the coupling constant becomes strong, or the radius of the
circle becomes small in the $R\to\infty$ limit. Thus $R\to\infty$ does
not correspond to a simple limit in these theories. However, for the
$E_8\times E_8$ theory, the $R\to\infty$ limit corresponds to large radius
limit of the theory at {\it fixed value} of the $E_8\times E_8$ coupling
constant. Thus in this limit the theory reduces to ten dimensional
$E_8\times E_8$ heterotic string theory.

Recall that we are sitting at the point in the moduli space where
$E_8\times E_8$ is broken down to SO(16)$\times$SO(16). Thus we have a 
constant
background gauge field configuration. 
In order to break $E_8\times E_8$
to SO(16)$\times$SO(16) we need a fixed amount of Wilson line, {\it i.e.}
a fixed amount of $A^I_9\cdot r_{E_8\times E_8}$, where $A^I_\mu$
($1\le I\le 16$) denote
the gauge fields in the Cartan subalgebra of $E_8\times E_8$.\footnote{We
are using a coordinate system in which $G^{(E_8\times E_8)}_{99}=1$, where
$x^9$ is the coordinate on the circle.} Since in
the limit of large $R$, $r_{E_8\times E_8}$ also becomes large, we
get $A^I_9\to 0$. In other words in this limit the background gauge field
goes to zero, and locally we recover the $E_8\times E_8$ heterotic string
theory without any background gauge field.

This establishes the duality between $M$-theory on $S^1/Z_2$ and the
$E_8\times E_8$ heterotic string theory.\footnote{In order to avoid
misuse of this procedure, I would like to emphasize again that for this
procedure to work we must ensure that the large radius limit on the two
sides match. As an example let us consider the duality between $M$-theory
on $(K3\times S^1)/Z_2$\cite{MTHEORY} and the Dabholkar-Park 
orientifold\cite{DABPAR} that can be described as type IIB on $K3/Z_2'$.
By compactifying the first theory on another circle $(S^1)'$ we get type
IIA on $(K3\times S^1)/Z_2$. An $R\to (1/R)$ duality transformation on
$S^1$ converts this to type IIB on $(K3\times (S^1)'')/Z_2''$. It is easy
to verify that this $Z_2''$ does not act on $(S^1)''$, and in fact has
the same action as the $Z_2'$ used in the construction of ref.\cite{DABPAR}.
{}From this one might be tempted to conclude that this `proves' the duality
between $M$-theory on $(K3\times S^1)/Z_2$ and type IIB on $K3/Z_2''$.
However, by following the duality relations carefully one finds that the
limit where the radius of $(S^1)'$ goes to infinity keeping the volume of
$K3$ and $S^1$ measured in the $M$-theory metric fixed, does not correspond
to taking the radius of $(S^1)''$ to infinity keeping the volume of $K3$
measured in the type IIB metric and the type IIB coupling constant fixed.
Thus this analysis {\it does not} prove the duality between these two 
theories, although there is a more involved argument\cite{MORBI} which
does establish this duality.}
By carefully following this
argument we can also identify the origin of the gauge
symmetry in $M$-theory. For this we note that in identifying $M$-theory
on $S^1/Z_2$ with a particular limit of type IIA theory on $S^1/Z_2$, we
have identified the manifolds $S^1/Z_2$ in the two theories. In particular
the two boundaries of $S^1/Z_2$ in the 
$M$-theory compactification gets mapped
to the two boundaries of $S^1/Z_2$ in the type IIA compactification. Since
the two SO(16) factors in the type IIA compactification have their origin
at the two boundaries of $S^1/Z_2$, the same must be the case in $M$-theory.
Enhancement of SO(16) to $E_8$ is a non-perturbative phenomenon from the
type IIA / $M$-theory point of view that cannot 
be understood directly in these
theories in a simple manner.

Finally, I would like to mention that
the duality between $M$-theory 
on $T^5/Z_2$ and type IIB string theory on K3, discussed in the previous
subsection, can also be established this way.

\sectiono{Compactification involving $F$-theory} \label{s5}

In conventional compactification of type IIB theory, the complex field
$\lambda_{IIB}$ is constant on the internal manifold. $F$-theory is
a way of compactifying type IIB theory which does not suffer from this
constraint\cite{VAFAF}. The starting point in an $F$-theory
compactification is an elliptically fibered manifold $\MM$ which
is constructed by erecting at every point on a base manifold
$B$ a copy of the two dimensional torus $T^2$, with the
moduli of $T^2$ varying over the base in general. Thus if $\vec z$ denotes
the coordinate on the base $B$ and $\tau$ denotes the 
complex structure modulus
of the torus, then for a given $\MM$ we have a function $\tau(\vec z)$
specifying the variation of $T^2$ on $B$. By definition\cite{VAFAF}, 
$F$-theory on $\MM$ is type IIB string theory compactified on $B$, with,
\be \label{el1}
\lambda_{IIB}(\vec z)=\tau(\vec z)\, .
\ee
An example of such a manifold
is elliptically fibered K3, which can be viewed
as $T^2$ fibered over $CP^1$ with appropriate $\tau(z)$.
Thus $F$-theory compactification on this manifold corresponds to type IIB 
compactification on $CP^1$ with appropriately varying $\lambda_{IIB}(z)$.
This theory was conjectured to be dual to heterotic string theory
compactified on $T^2$\cite{VAFAF}. By applying this duality conjecture
fiberwise on Calabi-Yau manifolds admitting K3 fibration, many new
duality conjectures relating $F$-theory compactification on Calabi-Yau
manifolds and different heterotic compactifications have been 
`derived'\cite{VAFMOR}.

We shall now see how the original duality conjecture between heterotic 
string theory on $T^2$ and $F$-theory on elliptically fibered K3
manifold can be `derived' using the methods of section \ref{s2}. 
Since this has
been discussed in detail in ref.\cite{FTHEORY} our discussion will be
very brief. We go to the $T^4/\II_4$ orbifold limit of K3 where $\II_4$
denotes the reversal of sign of all four coordinates on the torus. 
Reexpressing $T^4/\II_4$ as $(T^2\times (T^2)')/\II_2\cdot\II_2'$, we
see that $T^4/\II_4$ can be regarded as $T^2$ fibered over $(T^2)'/\II_2'$,
with a twist $\II_2$ on the fiber as we move along a closed
cycle around a fixed point
on $(T^2)'/\II_2'$. By definition $F$-theory compactified on $T^4/\II_4$
can then be regarded as type IIB theory compactified on 
$(T^2)'/\sigma\cdot\II_2'$ 
where $\sigma$ represents the SL(2,Z) transformation 
$\pmatrix{-1 & \cr & -1}$. By
studying the action of this SL(2,Z) transformation on the massless fields
in the theory given in eq.\refb{e1} one can identify this transformation to 
$(-1)^{F_L}\cdot\Omega$. Thus $F$-theory on $T^4/\II_4$ can be identified
to type IIB on $(T^2)'/(-1)^{F_L}\cdot\Omega\cdot\II_2'$. As was shown in 
section \ref{s3.2} this theory in turn is dual to heterotic string theory
on $T^2$. Thus we see that $F$-theory on K3 in the orbifold limit is dual
to heterotic string theory on $T^2$. Once the duality is established at
one point in the moduli space, it is guaranteed to hold at all other points.
This has been discussed explicitly in ref.\cite{FTHEORY}

This procedure of `deriving' duality conjectures 
involving $F$-theory has been
used to `derive' many other duality conjectures involving 
$F$-theory and orientifolds\cite{GIMJOH,BLUZAF,DABPAR3,GOPMUK}. 
Indeed, all conjectured dualities
involving $F$-theory can be `derived' either using this procedure, or by 
fiberwise application of the duality 
between $F$-theory on elliptically fibered
K3 and heterotic string theory on $T^2$.

\section{Conclusion}

To summarise, we have shown that all conjectured dualities involving string,
$M$- and $F$-
theories and their compactifications
can be `derived' from the
duality between type I and SO(32) heterotic string theories in
ten dimensions and the definitions of $M$- and $F$- theories.
The set of rules that we have followed during these `derivations' are 
outlined in section \ref{s2}. Some of them are intuitively more obvious
than others. The least obvious of these rules is the fiberwise
application of duality discussed in section \ref{s2.1}. 
This is mainly based on the assumption that if we can argue equivalence
between two string compactifications in the bulk of the compact manifold,
then it automatically forces the two theories to be the same on 
boundaries / singular submanifolds of codimension $\ge 1$ where there
is no direct argument for the equivalence between the two theories. Perhaps
the success of this procedure itself teaches us something deep about
non-perturbative string theory which we have not yet been able to uncover.

\noindent{\bf Acknowledgement}:
I wish to thank the organisers of the conference for providing a very 
stimulating environment. 
I would like to thank J. Blum, K. Dasgupta, R. Gopakumar, A. Hanany, C. Hull,
K. Intrilligator, S. Mukhi, M. Ruiz-Altaba,
C. Vafa and E. Witten for useful discussions at various stages of this work. 

\appendix

\sectiono{Minimal set of T-duality conjectures} \label{s3.3}

In our analysis we have used T-duality symmetries of
various string theories indiscriminately. We shall now try to identify the
minimal set of T-duality conjectures from which all T-duality symmetries
can be derived following the principles outlined in section \ref{s2}.

We shall begin with type II theories. 
Upon compactification on $S^1$, type IIA
and type IIB theories transform into each other under the $R\to(1/R)$
duality transformations\cite{IIAIIB,POLET}. 
This duality cannot be attributed to any gauge
or general coordinate transformation and must be added to the list of
input conjectures. Let us now consider type IIA and type IIB theories
compactified on $T^2$. Since these two theories compactified
on $S^1$ are dual to each other, 
they will be dual to each other upon compactification on $T^2$ as well.
But besides this,
both these theories have SL(2,Z)$\times$SL(2,Z)$'$ self-duality
symmetries. In the type IIA theory SL(2,Z) can be identified to
the global diffeomorphism symmetry of the torus and hence is part of the
general coordinate transformation of the ten dimensional type IIA theory.
SL(2,Z)$'$ on the other hand acts as modular transformation on
the (complexified) Kahler modulus of the torus 
and has no simple geometric interpretation. In type IIB theory their roles
get reversed.  Now SL(2,Z)$'$ can be identified to
the global diffeomorphism symmetry of the torus and hence is part of the
general coordinate transformation of the ten dimensional type IIB theory.
SL(2,Z) acts as modular transformation on
the (complexified) Kahler modulus of the torus 
and has no simple geometric interpretation. Thus if we believe in 
general coordinate invariance of type IIA and type IIB theories, then
the full T-duality symmetries of type IIA/IIB theories on $T^2$ follow as
a consequence of the duality between type IIA and type IIB theories
compactified on $S^1$. It can be shown that all the T-duality symmetries
of type IIA/IIB theories compactified on $T^n$ can be derived in a 
similar manner.  

Let us now turn to T-dualities involving heterotic string theory.
We start in nine dimensions by compactifying both the heterotic
string theories on $S^1$.
In this case there is an $R\to (1/R)$ duality transformation that 
relates the SO(32) heterotic string theory with gauge group broken to
SO(16)$\times$SO(16) via Wilson line to $E_8\times E_8$ heterotic
string theory with gauge group broken to SO(16)$\times$SO(16) via Wilson
line. Furthermore, in the absence of Wilson lines each theory has
an $R\to(1/R)$ self-duality symmetry; this transformation gets modified
in the presence of Wilson lines.
However, it can be shown that the $R\to(1/R)$ self-duality in the
$E_8\times E_8$ heterotic string theory gets mapped to an SO(32) gauge
transformation in the SO(32) heterotic string theory under the map that
relates the two heterotic string theories on $S^1$. This is done as
follows. While viewing the nine dimensional theory as a compactification
of $E_8\times E_8$ (SO(32)) heterotic string theory, it is natural to
regard the O(17,1) lattice as $\Gamma_8\oplus\Gamma_8\oplus\Gamma_{1,1}$
($\Gamma_{16}\oplus\Gamma'_{1,1}$). The O(17,1) rotation that relates
these two lattices was constructed explicitly in ref.\cite{SOE8}. Now, the
$R\to (1/R)$ self-duality of the $E_8\times E_8$ theory exchanges the
two basis vectors of $\Gamma_{1,1}$. By using the O(17,1) rotation 
constructed in ref.\cite{SOE8} one can show that exchanging the two basis
vectors of $\Gamma_{1,1}$ corresponds to a transformation that leaves
$\Gamma'_{1,1}$ invariant and acts only on $\Gamma_{16}$. Thus it must
correspond to a gauge transformation in SO(32). Exactly similar analysis
shows that the $R\to (1/R)$ self-duality of 
the SO(32) heterotic string theory
compactified on $S^1$ gets mapped to a gauge transformation in the
$E_8\times E_8$ theory. Hence gauge invariance of the two heterotic
string theories, together with the duality between them upon compactification
on $S^1$, automatically implies self-duality of both of them on $S^1$.
This result is also supported by the fact that this self-duality can be
regarded as part of
the SU(2) enhanced gauge symmetry that 
appears at the self-dual point\cite{IIAIIB}.

Compactification of either of these heterotic string theories on $T^n$
has a T-duality group $O(16+n,n;Z)$, which can be similarly
derived by combining
the duality symmetries of these theories on $S^1$, and gauge
and general coordinate invariance. Thus the only T-duality that needs
to be added to the list of independent duality conjectures is the
duality between SO(32) heterotic and $E_8\times E_8$ heterotic string
theories on $S^1$.

It has also been argued recently that all
mirror symmetries involving compactification on Calabi-Yau and K3 
manifolds can be understood as a result of fiberwise application of
$R\to(1/R)$ duality transformations\cite{YAUSTR,MORR}. Thus these
can also be
traced to the duality between the two type II and the two heterotic
string theories compactified on $S^1$.

\end{document}